\documentclass[twocolumn,prb]{revtex4-2}
\usepackage{titlesec}
\usepackage{subfigure}
\usepackage{amsmath}
\usepackage{amsfonts}
\usepackage{amssymb}
\usepackage{feynmp}
\usepackage{graphicx}
\usepackage{setspace}
\usepackage{comment}
\usepackage{dsfont}
\usepackage{color}
\usepackage[pdftex,colorlinks=true,linkcolor=blue,citecolor=blue]{hyperref}

\begin{document}

\title{Topological inverse Anderson insulator}

\author{Zheng-Wei Zuo}
\author{Jing-Run Lin}
\author{Dawei Kang}

\affiliation{School of Physics and Engineering, Henan University of Science and Technology, Luoyang 471023, China}
\date{\today}

\begin{abstract}
A different type of topological phase dubbed topological inverse Anderson insulators is proposed, which is characterized by the disorder-induced extended bulk states from the flat-band localization and topological edge states. Based on the topological invariant, the behaviors of the localization length of the zero-energy modes, and quantum transport, we identify its existence in several all-band-flat models with the disordered potentials or hopping including the $\pi$-flux Creutz ladder, the fully dimerized Su-Schrieffer-Heeger chain, and $\pi$-flux diamond chain. Unlike the topological Anderson insulator, where disorder induces localization and exponential suppression of transport, the disorder-assisted quantum ballistic coherent transport can appear in the topological inverse Anderson insulator. In addition, our proposal and results could be realized by the current experimental techniques.  
\end{abstract}

\maketitle

\section{Introduction}
Quantum matter with single-electron flat bands\cite{TasakiH98RTP, LiuZ14CPB, LeykamD18APX, CalugaruD21NTP, RegnaultN22NT} has become an ideal quantum platform to investigate the various strongly correlated electronic states. In particular, Moir\'e superlattice systems such as magic-angle twisted bilayer graphene\cite{CaoY18NT1, CaoY18NT2} with topological flat bands have been intensively studied, where the fractional Chern insulators have been experimentally demonstrated recently\cite{CaiJQ23NT,ZengYH23NT,ParkH23NT,XuF23PRX}. Due to the quenched kinetic energy in the complete flat bands, the wave group velocity of the electrons is strictly zero for all momenta in the Brillouin zone. The eigenstates of these dispersionless flat bands are characterized by the compact localized states, which are sharply localized within a small finite number of lattice sites. The compact localized states are caused by destructive interference and the local spatial symmetries. When the magnetic field is incorporated, the complete flat-band localization phenomenon called Aharonov-Bohm caging\cite{VidalJ98PRL} appears, where the eigenstates are completely localized due to Aharonov-Bohm destructive interference. When the electron-electron interaction is further considered, the various many-body phases can emerge such as ferromagnetism\cite{MielkeA91JPA, MielkeA91JPA2, MielkeA92JPA, Tasaki92PRL}, Wigner crystal\cite{WuCJ07PRL, WuCJ08PRB}, superconductivity\cite{MiyaharaS07PC, Volovik18JETP}, fractional Chern insulators\cite{Bergholtz13, ParameswaranSA13} and quantum many-body scars\cite{KunoY20PRB, KunoY21PRB, NicolauE23PRB, NicolauE23PRB2}.

According to the band-crossing singularity of Bloch states, the flat bands can be classified into two classes: singular and non-singular flat bands\cite{BergmanDL08PRB, RhimJW19PRB, RhimJW20NT}. In the singular flat band, the compact localized states do not form a complete set spanning the singular flat band and the non-contractible loop states with the robust boundary modes exhibiting the nontrivial real-space topology can appear. The singular flat band becomes a new platform to investigate the geometrical (curvature and metric) properties of Bloch states. On the other hand, the disordered flat bands displays rich quantum behaviors including inverse Anderson transitions\cite{GodaM06PRL, LonghiS21OL, LiH22PRL, WangHT22PRB, ZhangWX23PRL}, multifractality\cite{ChalkerJT10PRB}, and localization lengths diverging with unconventional exponents\cite{LeykamD13PRB, FlachS14EPL, BodyfeltJD14PRL}.

The topological phases of matter with exotic bulk phenomena and robust boundary effects have become a surging field in condensed matter physics\cite{Bernevig13Book, MoessnerR21Book, Hasan10RMP, QiXL11RMP, Franz15RMP, WenXG17RMP, BergholtzEJ21RMP}. Disorders and impurities are ubiquitous in real quantum matter, playing an important role in the appearance of the different quantum phases. The interplay of disorder and topological states is extensively explored. For example, strong disorder can induce the trivial systems into topological Anderson insulator phases\cite{LiJ09PRL, GrothCW09PRL, GuoHM10PRL, XingYX11PRB, MondragonShem14PRL, AltlandA14PRL2}, which have been the subject of extensive experimental and theoretical studies\cite{MeierEJ18SCI, StuetzerS18NT, LiuGG20PRL, CuiXH22PRL, ChenR23PRB, ChengXY23PRB, LoioH24PRB}. When the quasiperiodic potentials and/or spatially correlated disorders are added, various remarkable features such as reentrant localization-delocalization transition appear\cite{ShilpiR21PRL, ZuoZW22PRA, HanWQ22PRB, GoncalvesM23PRB, QiR23PRB, WangHY23PRB, PadhanvA24PRB}. By applying a magnetic field or introducing various symmetries, the different types of topological flat bands could be constructed\cite{Bergholtz13, ParameswaranSA13, Creutz99PRL, KremerM20NTC, BerciouxD17AP}.

The interplay of disorder and topological flat bands is an intriguing subject. In this paper, we theoretically investigate the complex competition of the Anderson localization and the flat-band localization in several all-band-flat models and find a different kind of topological phase termed topological inverse Anderson insulator, where the disorders induce the localized bulk states in the dispersionless flat bands into extended states and the topological edge states are located at the two ends. We provide a detailed demonstration of this topological phase by the disordered $\pi$-flux Creutz ladder, fully dimerized Su-Schrieffer-Heeger (SSH) chain and the $\pi$-flux diamond chain using the topological invariant, the behaviors of localization length of the zero-energy modes, quantum transport. By tuning the disorder strength, we show that the system undergoes a phase transition from the topological insulator with flat-band localization to the topological inverse Anderson insulator. The extended bulk states are demonstrated by rigorous analytic proof and quantum transport. The disorder-assisted quantum ballistic coherent transport becomes possible in this topological phase. This work opens  another direction in the search for topological quantum matter, where the disordered systems are topological insulators with the extended bulk state induced by disorder and topological edge states.

This paper is organized as follows: Sec. \ref{CreutzModel} discusses the topological features and localization properties of the $\pi$-flux Creutz ladder with the antisymmetric-correlated disorder. The phase diagram is given in Sec. \ref{PhaseDiagram}. Then, we investigate the quantum transport characterization of the disordered $\pi$-flux Creutz ladder in Sec. \ref{Transport}. Sec. \ref{Diamond} analyzes the topological and transport properties of the $\pi$-flux diamond chain with the antisymmetric-correlated disorder. Conclusions and discussions are presented in Sec. \ref{Conclusions}.

\section{Model}
\subsection{Creutz Ladder}\label{CreutzModel}
\begin{figure}[tb]
\centering
\includegraphics[width=\columnwidth]{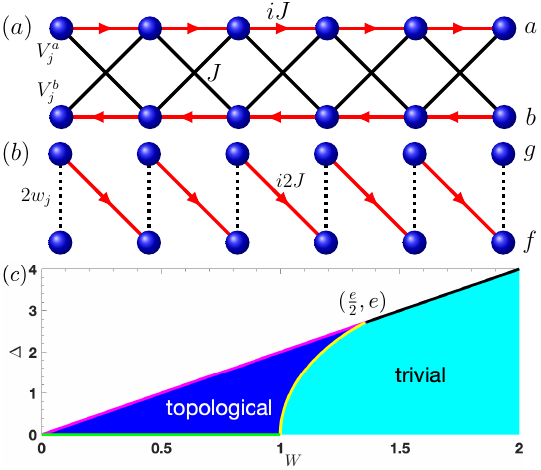}
\caption{$(a)$ Sketch of the disordered $\pi$-flux Creutz ladder with on-site disorders $V^a_j$ and $V^b_j$. Blue balls denote the lattice sites. The arrows depict the sign of the intra-chain hopping $iJ$. The cross-link hoppings are $J$. $(b)$ Mapping of the disordered $\pi$-flux Creutz ladder to a disordered SSH chain. $(c)$ The phase diagram for the disordered $\pi$-flux Creutz ladder with $J=1$, and $N=10^4$ case.}
\label{Fig1}
\end{figure}

Our starting point is the disordered $\pi$-flux Creutz ladder \cite{Creutz99PRL}(illustrated in Fig.\ref{Fig1}a), which is described by the real-space Hamiltonian
\begin{align}
H_0=&-\sum_j^{N-1}[iJ(a_{j+1}^{\dagger}a_j-b_{j+1}^{\dagger}b_j)+J(a_{j+1}^{\dagger}b_j+b_{j+1}^{\dagger}a_j) \notag \\ 
&+H.c.]+\sum_j^N(V^a_ja_j^{\dagger}a_j+V^b_j b_j^{\dagger}b_j),\label{H0}
\end{align}
in which $a_{j}^{\dagger}$ and $b_{j}^{\dagger}$ ($a_{j}$ and $b_{j}$) are fermionic creation (annihilation) operators on the upper (denoted by $a$) and lower (denoted by $b$) chains, respectively. $J$ is the parameter controlling the strengths of the intra-chain and inter-chain couplings. $V^a_j$ and $V^b_j$ are the on-site disorder potentials at the chains $a$ and $b$. Here, we consider the antisymmetric-correlated disorder $V^a_j=-V^b_j$ case. $N$ is the number of unit cells.

In the clean limit, the ladder possesses two complete flat bands, $E_\pm =\pm2J$ due to the destructive interference of the hoppings. The compact localized eigenstates of the two highly degenerate flat bands are given by $\frac{1}{2}[ia_{j+1}^{\dagger}+b_{j+1}^{\dagger}\pm a_j^{\dagger}\pm ib_j^{\dagger}]|0\rangle$ ($|0\rangle$ indicates the vacuum state). Consequently, when a particle is inserted in the system, it is localized on the four neighbor sites, known as the Aharonov-Bohm caging phenomenon. The two flat bands have the disorder-free localization and topological properties with the Berry phase $\pi$ (winding number $\pm 1$)\cite{Creutz99PRL, JunemannJ17PRX, ZuritaJ19AQT, KunoY20PRA, KunoY20NJP, YoshihitoK20PRB, OritoT20PRB}. For a finite system under an open boundary condition (OBC), two topological zero-energy edge states located at the ends appear.

To gain insight into the features of this disordered $\pi$-flux Creutz ladder system, it is convenient to write the Hamiltonian Eq.\ref{H0} in a different basis. By performing the unitary transformation of the following operators and defining $w_j=(V^b_j-V^a_j)/2$, we use
\begin{equation}
\left(\begin{array}{c}
f_j^{\dagger} \\
g_j^{\dagger}
\end{array}\right)=\frac{1}{\sqrt{2}}\left(\begin{array}{cc}
i & 1 \\
-i & 1
\end{array}\right)\left(\begin{array}{c}
a_j^{\dagger} \\
b_j^{\dagger}
\end{array}\right),
\end{equation}
and the Hamiltonian of this disordered $\pi$-flux Creutz ladder is cast into the form
\begin{equation} 
H_{SSH}=\sum_j^{N-1}(i2Jf_{j+1}^{\dagger}g_j+H.c.)+\sum_j^{N}2w_j(f_j^{\dagger}g_j+g_j^{\dagger}f_j). \label{SSH}
\end{equation}

It is easy to see that the disordered $\pi$-flux Creutz ladder is mapped into a SSH chain with inter-cell coupling $i2J$ and intra-cell disordered-coupling $2w_{j}$, as shown in Fig.\ref{Fig1}b. Because of the antisymmetric-correlated disorder, the chiral symmetry of the $\pi$-flux Creutz ladder is explicitly broken. By way of the operator transformation, the mapped disordered SSH chain has chiral symmetry, which indicates the existence of a hidden chiral symmetry in the disordered $\pi$-flux Creutz ladder. In the clean limit, for the disordered SSH chain under periodic boundary condition, the system becomes fully dimerized and has two complete flat bands, which is the same as the above discussion of the clean $\pi$-flux Creutz ladder. The SSH chain falls apart to dimers and the states become localized. Thus, an electron in the bulk cannot move along the chain. 

Although the disordered SSH chain has been the subject of extensive studies, here we provide a deeper understanding of the topological features. From the Hamiltonian $H_{SSH}$ of the disordered SSH chain, one can see that as long as the intercell hopping $w_{j}\neq 0$, the disorder-assisted quantum transport is allowed. Therefore, the flat-band localization could be destroyed by the antisymmetric-correlated disorder. This implies that the bulk states of the disordered Creutz ladder can be delocalized. 

More specifically, the probability density function $f(w)$ of the disorder potential $w_{j}$ is chosen as\cite{LonghiS21OL}
\begin{equation}
f(w)=\left\{\begin{array}{c}
\frac{1}{2 \Delta},|w_j \pm W|<\frac{\Delta}{2} \\
0, \text { otherwise }
\end{array}\right.
\end{equation}
where $w_j$ is an independent stochastic variable with the same probability density function of zero mean, $W$ is the disorder potential strength and $\Delta$ is the uniformly distributed width of the disorder potential in the range $(\pm W-\frac{\Delta}{2},\pm W+\frac{\Delta}{2})$. In the limit $\Delta\rightarrow2W$, $f(w)$\ is uniformly distributed in the range ($-2W,2W$). In the other limit $\Delta\rightarrow0$, $f(w) $ is the Bernoulli distribution (binary disorder) and $w_{j}$ takes only the two values $\pm W$ with the same probability.

First, we investigate electronic properties of the SSH chain with the Bernoulli distribution disordered-hopping case. Supposing that the eigenstates of one-particle are given by $\left|\psi_n\right\rangle=\sum_j(\psi^f_{j, n} f_j^{\dagger}+\psi^g_{j, n} g_j^{\dagger})|0\rangle$, we can solve the Hamiltonian Eq.\ref{SSH} to obtain the difference equations
\begin{align}
E\psi^f_{j, n}&=i2J\psi^g_{j-1, n}+2We^{i\pi\delta_j}\psi^g_{j, n}, \label{Diff1} \\ 
E\psi^g_{j, n}&=-i2J\psi^f_{j+1, n}+2We^{i\pi\delta_j}\psi^f_{j, n}, \label{Diff2}
\end{align}
where the phases $e^{i\pi\delta_j}$ ($\delta_j=0,1$) depend on the disorder values $w_j=\pm W$. Now we introduce the gauge transformation $\psi^f_{j, n}=\tilde{\psi}^f_{j, n}\exp(-i\pi\sum_{l}^{j-1}\delta_l)$ and $\psi^g_{j, n}=\tilde{\psi}^g_{j, n}\exp(-i\pi\sum_{l}^{j-1}\delta_l)$. Then, these phases are gauged away and the difference equations take the following form
\begin{align}
E\tilde{\psi}^f_{j, n}&=i2J\tilde{\psi}^g_{j-1, n}+2W\tilde{\psi}^g_{j, n}, \\
E\tilde{\psi}^g_{j, n}&=-i2J\tilde{\psi}^f_{j+1, n}+2W\tilde{\psi}^f_{j, n}, 
\end{align}
which corresponds to the disorder-free conventional clean SSH chain with  inter-cell coupling $i2J$ and intra-cell coupling $2W$. So, the disorder-induced energy spectra of the disordered SSH chain (Creutz ladder) under a periodic boundary condition (PBC) become absolutely continuous and the two dispersive bands $E=\pm 2\sqrt{J^2+W^2+2JW\sin k}$, where $-\pi< k \leq \pi$ (lattice constant is set to unity) is the Bloch wavenumber. At the same time, the eigenstates become perfectly extended and of the Bloch type. Thus, the antisymmetric correlated disorder revives mobility in this flat-band system, which indicates an escape from Aharonov-Bohm caging and the destruction of the flat-band localization. This phenomenon is known as the inverse Anderson transition\cite{GodaM06PRL,LonghiS21OL,LiH22PRL,WangHT22PRB,ZhangWX23PRL}. When the coupling amplitudes $W<J$, the disordered SSH chain at $1/2$ filling is in the topological insulator phase with two zero-energy edge states.  It implies that for the antisymmetric-correlated disorder with Bernoulli distribution, the disordered $\pi$-flux Creutz ladder has disorder-induced extended bulk states of the Bloch type and two topological zero-energy edge states, which is a counterintuitive phenomenon. We dub such a disordered topological insulator with disorder-induced extended states and topological edge states the \emph{topological inverse Anderson insulator}. Due to this intriguing feature, we next investigate the phase diagram and the quantum transport properties of this disordered $\pi$-flux Creutz ladder. A comment is in order. Because the disordered $\pi$-flux Creutz ladder in the clean limit is in the topological phase, the topological phase transition does not take place when the antisymmetric correlated disorder is added (small disorder case). The system only undergoes the inverse Anderson transition.

\subsection{Phase diagram}
\label{PhaseDiagram}

As shown in Ref.\cite{MondragonShem14PRL}, the topological phase transition of the disordered SSH chain with chiral symmetry at $1/2$ filling is accompanied by the divergence of the localization length of the zero-energy modes. From Eq. \ref{SSH}, we can obtain the Lyapunov exponent $\lambda^{-1}$ (the reciprocal of the localization length) of the zero-energy modes
\begin{equation}
\lambda^{-1}=\left| \lim_{N\rightarrow\infty}\frac{1}{N}{\sum\limits_{j=1}^{N}} \ln\left| \frac{J}{w_j}\right| \right|=\left| \lim_{N\rightarrow\infty}\frac{1}{N}{\sum\limits_{j=1}^{N}} \ln\left| w_{j}\right| \right|,
\end{equation}
where we have set $J=1$ for convenience. According to the probability density function $f(w)$ of disorder and Birkhoff's ergodic theorem, we can use the ensemble average to evaluate the Lyapunov exponent as
\begin{align}
\lambda^{-1} =2\left( \ln\frac{\left| \frac{2W+\Delta}{2}\right| ^{W/\Delta+1/2}}{\left| \frac{2W-\Delta}{2}\right| ^{W/\Delta-1/2}}-\Delta\right). \label{Lyapunov}
\end{align}

Therefore, the topological phase boundary is identified by the exact solution of the following equation as follows:
\begin{equation}
(W+\frac{\Delta}{2})\ln\left| W+\frac{\Delta}{2}\right| -(W-\frac{\Delta}{2})\ln\left| W-\frac{\Delta}{2}\right| =\Delta. \label{Boundary}
\end{equation}

On the other hand, the chiral symmetry of the disordered SSH chain (Eq.\ref{SSH}) is maintained, the real space winding number and the electric polarization\cite{MondragonShem14PRL} can be used to recognize the topological phases. Here we use another topological invariant $\mathcal{Q}$\cite{FulgaIC11PRB,LonghiS20OL} to characterize the topological states. The topological invariant $\mathcal{Q}$ is constructed from the reflection matrix and counts the number of stable bound states at the ends and can be applied to the larger disordered system, which reads
\begin{equation}
\mathcal{Q}=\frac{1}{2}\left(1-\operatorname{sign} \left[\prod_j w_j^2-\prod_j J^2 \right]\right).
\end{equation}

In the large $N$ limit, we can write $\prod_j^N w_j^2=V^{2N}$, where $\ln V=\lim_{N\rightarrow\infty}\frac{1}{N}\sum\limits_{j=1}^{N}\ln\left| w_j\right|$. Compared with Eq.\ref{Lyapunov}, it is easy to see that $\ln V$ is exactly the Lyapunov exponent except the sign function ($\left| \ln V\right|=\lambda^{-1}$). In consequence, the divergence points of the localization length correspond to phase transition of the topological invariant $\mathcal{Q}$. 

According to the localization length and the topological invariant $\mathcal{Q}$, the phase diagram of the disordered $\pi$-flux Creutz ladder is shown in Fig.\ref{Fig1}c.  The analytic boundary for the divergence of the localization length (Eq.\ref{Lyapunov}) is depicted by the yellow solid curve in Fig.\ref{Fig1}c. Thus, our numerical calculations demonstrate that the critical phase transition points of the divergence of the localization length and topological invariant $\mathcal{Q}$ coincide with each other. The green line denotes the topological inverse Anderson insulator under the Bernoulli distribution disorder and the pink line represents the topological Anderson insulator\cite{MondragonShem14PRL} under the uniformly distributed disorder ($\Delta=2W$) in Fig.\ref{Fig1}c. In the blue region, the system is in the topological insulator phase with two zero-energy modes. While in the cyan region (strong disorder case), the system becomes topologically trivial. For a fixed disorder strength $W$, as the disorder width $\Delta$ increases from zero, the extended bulk states become more and more localized and the system evolves from the topological inverse Anderson insulator to conventional topological Anderson insulator phases.

\begin{figure}[t]
\centering
\includegraphics[width=\columnwidth]{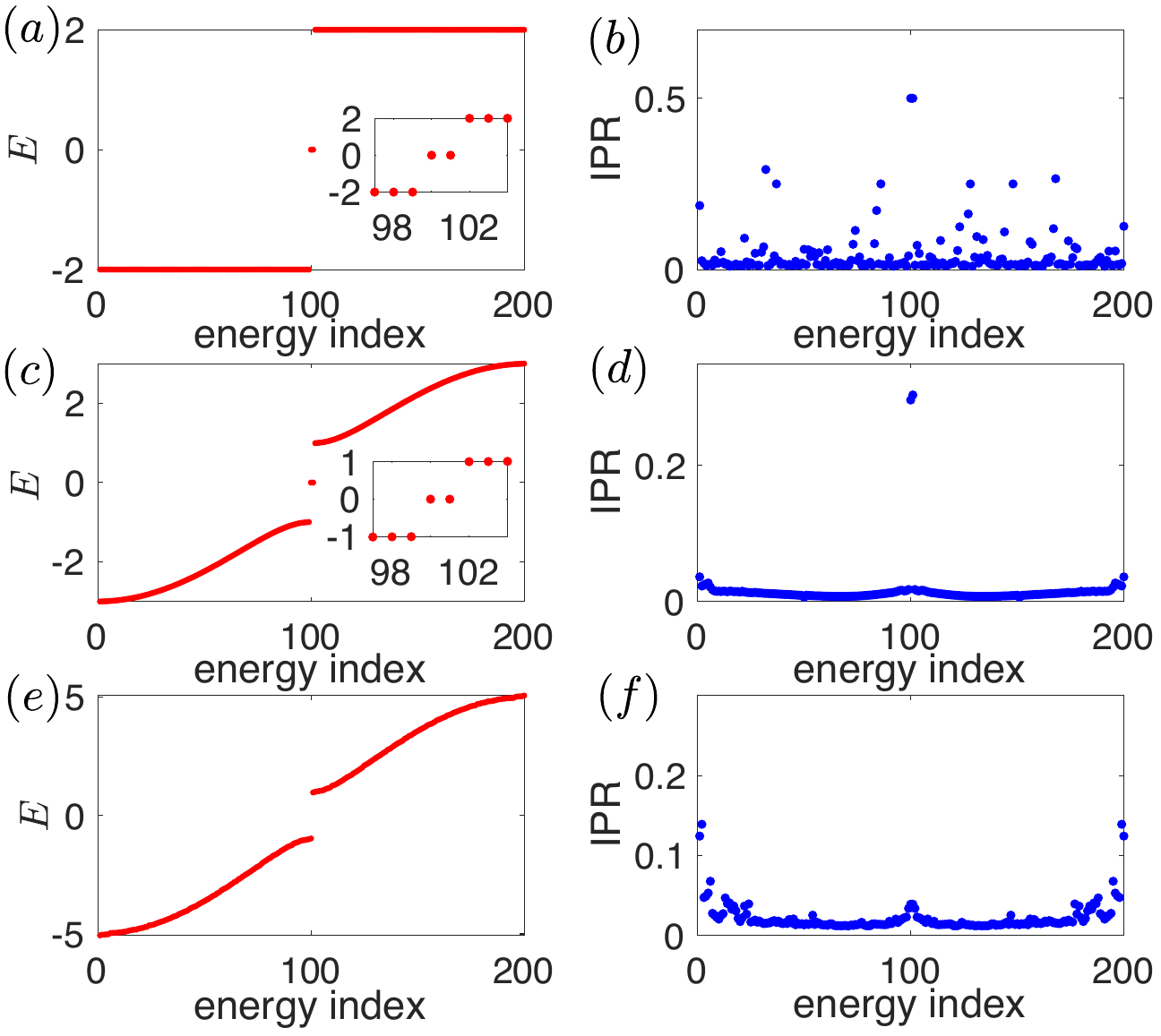}
\caption{The energies (red dots) and IPR (blue dots) for the disordered $\pi$-flux Creutz ladder with $N=100$, and $J=1$ under the OBC. $(a, b)$ In the clean limit case $W=\Delta=0$. The zoomed inset illustrates the two topological zero-energy edge modes. $(c, d)$ $W=0.5$, and $\Delta=0.02$ with one disorder configuration. $(e, f)$ $W=1.5$, and $\Delta=0.2$ with one disorder configuration.}
\label{Fig2}
\end{figure}

\begin{figure}[tp]
\centering
\includegraphics[width=\columnwidth]{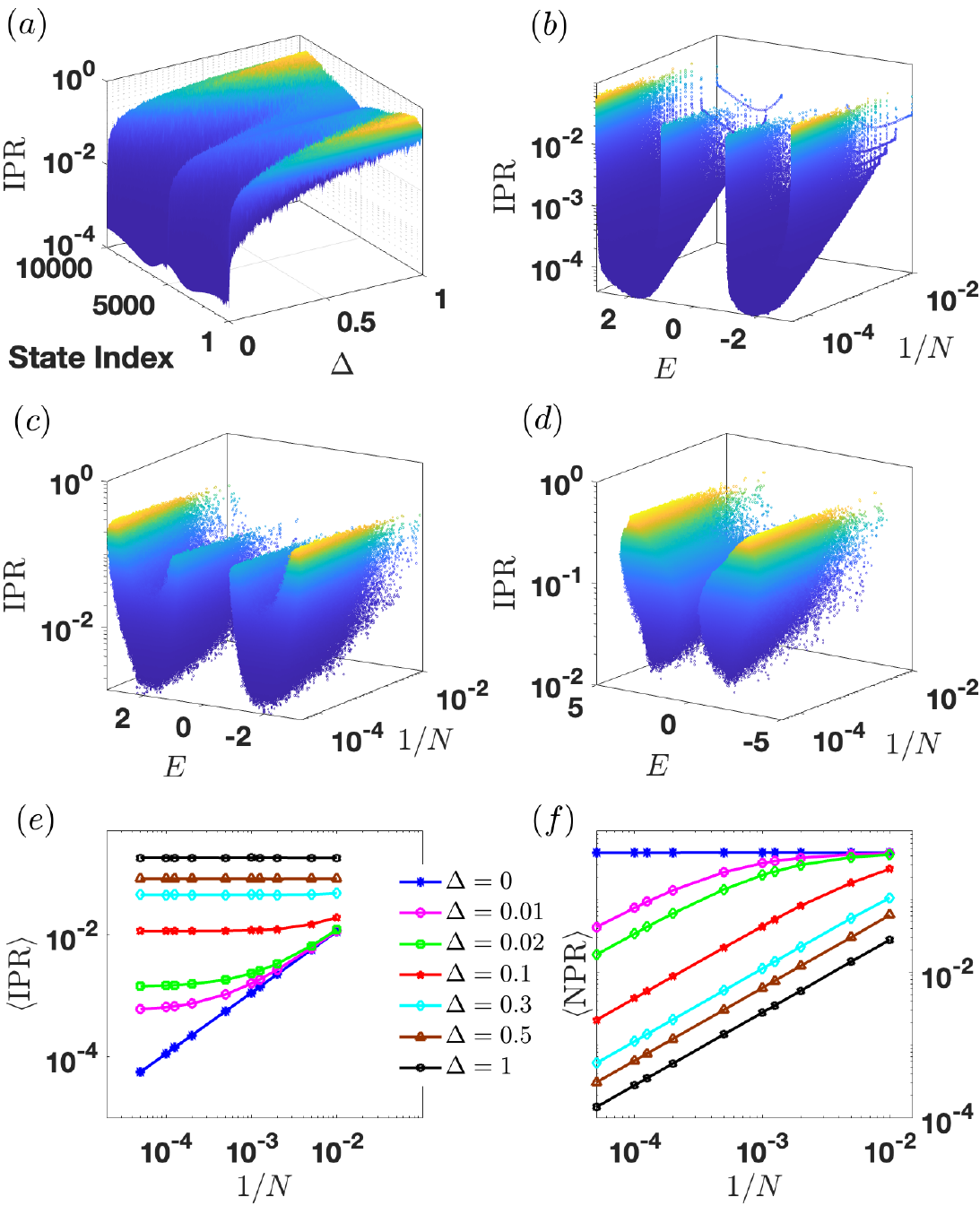}
\caption{$(a)$ The IPR values for the disordered $\pi$-flux Creutz ladder with $N=5000, J=1$, and $W=0.5$ as a function of $\Delta$ under one disorder configuration. $(b,c,d)$ The finite-size scaling analysis for IPR values with $J=1, W=0.5$, and $\Delta=0.02(b), 0.3(c), 0.9(d)$, respectively, under one disorder configuration. $(e, f)$ The finite-size scaling analysis of $\langle \mathrm{IPR}\rangle$  and $\langle \mathrm{NPR}\rangle$ with $W=0.5$ and averaged over 100 disorder realizations under the PBC.}
\label{Fig3}
\end{figure}

According to the bulk-edge correspondence, there are two topological edge bound states in topological insulator phases. To discern between localized and extended states, we use the inverse participation ratio (IPR) $\mathrm{IPR_{n}}=\sum_{j}^{N}\left| \psi_{n}(j)\right|^{4}$, where $\psi_{n}$ is the $n$-th normalized eigenstate of the system. For a perfectly extended state, the IPR scales as $1/N$ and vanishes in the thermodynamic limit, while it remains a finite value for a localized state. The energies (red dots) and IPR (blue dots) with the different strength disorder configurations are illustrated in Fig.\ref{Fig2} for the disordered $\pi$-flux Creutz ladder with $N=100$, and $J=1$ under OBC. In Fig.\ref{Fig2}(a-b), the clean $\pi$-flux Creutz ladder has two flat bands and two topological zero-energy edge states located at the two ends. When the antisymmetric-correlated disorder is switched on, the two flat bands become dispersive, as shown in Fig.\ref{Fig2}(c-d). The small IPR values of the bulk eigenstates indicate that they are extended states. At the same time, the two topological edge states located at the two ends are still fixed to zero-energy due to the hidden chiral symmetry and are robust against the disorder. As the disorder width $\Delta$ increases for a fixed disorder strength $W$, numerical calculations demonstrate that some IPR values become big, indicating the localized bulk states (see the bulk eigenstates of band-edge regions in Fig. \ref{Fig2}d, \ref{Fig2}f, and Fig.\ref{Fig3}a) and the coexistence of localized and delocalized bulk eigenstates. For the uniformly distributed case $\Delta=2W$, the bulk eigenstates become localized (Anderson insulator case). As the disorder strength $W$ further increases, the system enters into a trivial insulator phase without zero-energy states (see Fig.\ref{Fig2}(e-f)).

Next, we analyze in detail the coexistence properties and IPR behaviors of the localized and delocalized bulk eigenstates in the topological phase. Fig.\ref{Fig3}a shows the IPR values associated with the bulk eigenstates as a function of the disorder width $\Delta$, when the system parameters are $N=5000, J=1$, and $W=0.5$ under one disorder realization for each $\Delta$. It is easily to see that for a fixed finite size system, as the disorder width $\Delta$ increases, more and more bulk states become localized. Secondly, the finite-size scaling analysis for the IPR of the energy-dependent bulk states is carried out in Fig.\ref{Fig3}b-d, when the disorder strength $W=0.5$ and the disorder width $\Delta=(0.02, 0.3, 0.9)$ under one disorder realization. For the large-size systems, it is clearly shown that the bulk states of band-edge regions are localized, where the IPR values become finite values in the thermodynamic limit. On the other hand, the bulk states for the band-center regions are extended for the small $\Delta$ (up to $N=2\times 10^4$). As the disorder width $\Delta$ increases, we can see that the number of extended bulk states of band-center regions becomes more and more sparse. At the same time, the localization lengths of the bulk states become shorter and shorter. For the large-size systems, almost the bulk states become localized for larger $\Delta$. In addition, we illustrate the scaling analysis for the mean inverse participation ratio $\langle \mathrm{IPR}\rangle=\sum_{n}^{2N}\mathrm{IPR_{n}}/2N$ and the mean normalized participation ratio $\langle \mathrm{NPR}\rangle=\sum_{n}^{2N}\mathrm{NPR_{n}}/2N$ (where $\mathrm{NPR_{n}}=(2N \sum_{j}^{N}\left| \psi_{n}(j)\right|^{4})^{-1}$) with disorder strength $W=0.5$ and averaged over 100 disorder realizations in Fig.\ref{Fig3}e and Fig.\ref{Fig3}f, respectively. For the bulk extended phase ($\Delta=0$ case), the $\langle \mathrm{IPR}\rangle=0$ and $\langle \mathrm{NPR}\rangle \neq0$ in the thermodynamic limit. As the disorder width $\Delta$ increases, the system enters into the coexistence region characterized by nonzero values of $\langle \mathrm{IPR}\rangle$ and $\langle \mathrm{NPR}\rangle$. For all the localized phases ($\Delta=2W$, uniformly distributed disorder case), the $\langle \mathrm{IPR}\rangle \neq0$ and $\langle \mathrm{NPR}\rangle =0$, which implies that the bulk states are localized. The localization length for a fixed energy bulk eigenstates in the thermodynamic limit could be in principle calculated by the transfer matrix method\cite{MacKinnonA83ZPB, PendryJB92, SlevinK04PRB, SlevinK14NJP}.

\subsection{Transport properties}\label{Transport}
To further analyze the electronic properties of the topological inverse Anderson insulator, the experimental quantum mechanical time evolution of the electronic wave packet is used to investigate the characterization of the quantum transport. First, we put an electron in the center of chain $a$. The state vector of the electron at time $t=0$ is given by $\left|\psi_0\right\rangle= a_{N/2}^{\dagger}|0\rangle$. The mean square displacement of the electronic wave-packet can be calculated by $\left<\sigma(t)\right>=\sum_j^N j^2(|a_j(t)|^2+|b_j(t)|^2)$ as a function of time $t$ , where $a_j(t)$ and $b_j(t)$ are the occupation amplitudes of the dynamical evolution state $|\psi(t)\rangle=e^{-iH_0t}\left|\psi_0\right\rangle$ in the $j$-th unit cell.

\begin{figure}[tb]
\centering
\includegraphics[width=\columnwidth]{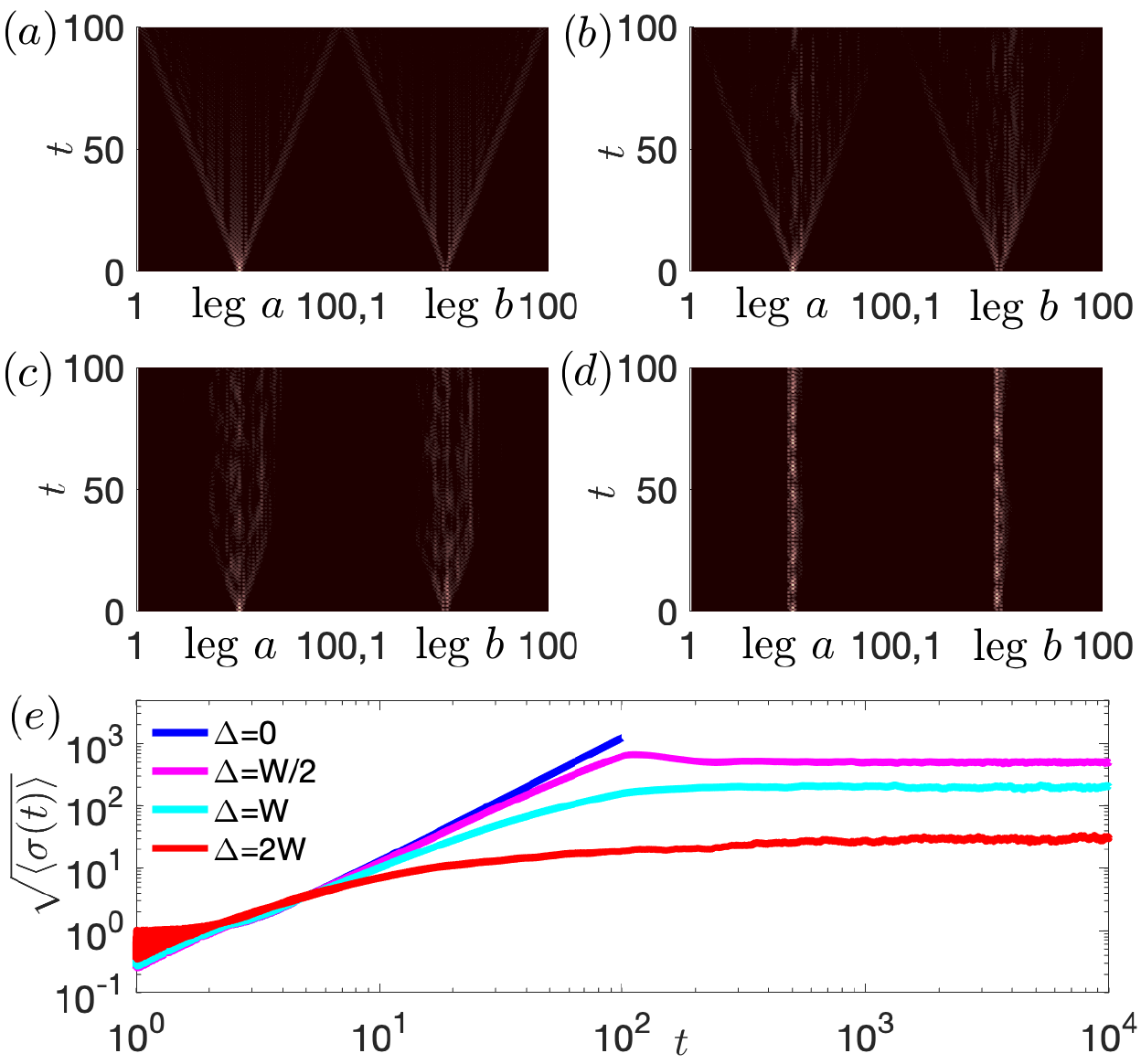}
\caption{The quantum mechanical time evolution of electronic wave-packet in the disordered $\pi$-flux Creutz ladder with $J=1$, $W=0.25$, and $N=100$ under the OBC. The electronic wave-packet is initially positioned on the center of leg $a$. $(a)$ $\Delta=0$. $(b)$ $\Delta=W/2$. $(c)$ $\Delta=W$. $(d)$ $\Delta=2W$. $(e)$ The log-log plot of the time evolution $\sqrt{\left<\sigma (t)\right>}$ averaged over 100 disorder realizations.}
\label{Fig4}
\end{figure}
Figure \ref{Fig4}(a-d) shows the time-dependent density distribution of the electronic wave-packet when the system parameters in the topological phase are $J=1$, $W=0.25$, and $N=100$ under OBC. One can observe that due to the different parameters of the probability density function for the disorder, the electronic wave packet displays different evolution dynamics. In consequence, the disorder-assisted quantum transport becomes possible. For the Bernoulli distribution disorder case (illustrated in Fig.\ref{Fig4}a), the ballistic coherent transport takes place. As shown in Ref.\cite{HiramotoH88JPSJ1, HiramotoH88JPSJ2}, the overall behavior of the diffusion width $\sqrt{\left<\sigma (t)\right>}$ at long times can be described by the power-law $\sqrt{\left<\sigma (t)\right>} \sim t^{\alpha}$. The index $\alpha$ continuously decreases from $1$ to $0$ when the eigenstates of the system change from extended to localized cases. For the Bernoulli disorder ($\Delta=0$) case, the bulk states are perfectly extended and of the Bloch type. Numeral calculations demonstrate that the diffusion linearly grows with time (make sure the wavefronts do not reach the boundaries), i.e. $\sqrt{\left<\sigma (t)\right>} \sim t$ ($\alpha=1$) indicating that the ballistic coherent transport appears, as marked by the blue curve in Fig.\ref{Fig4}e. We can also see that as the disorder width $\Delta$ increases, the diffusion width obeys the power-law behavior for small times (The oscillating behaviors of the diffusion width in the initial time are due to the different initial disorder configurations). As the time further increases (longer time) and $\Delta<1$, diffusion width deviates from the straight line and saturates a stable value, which indicates the existence of a finite localization length for the system. For the uniformly distributed disorder $\Delta=2W$, the diffusion width $\sqrt{\left<\sigma (t)\right>} \sim t^{\alpha}$ ($\alpha\rightarrow0$) highlighted by the red curve in Fig.\ref{Fig4}e converges to a stable constant, which shows that the wave diffusion is prevented and the most bulk eigenstates are localized (see Fig.\ref{Fig4}d). Thus, as the disorder width $\Delta$ increases, the electron motion undergoes the ballistic transport, superdiffusive transport, normal diffusive transport, subdiffusive transport, and finally localized transport, which consists of the IPR behaviors in Fig. \ref{Fig2}.

It should be noted that a conventional disordered SSH chain can emerge the topological inverse Anderson insulator phase from the trivial phase, if we choose the intra-cell hopping $J$ with the inter-cell hopping $w_j$. This type of disordered SSH chain with the Bernoulli distribution disordered hopping would be changed from a trivial fully dimerized insulator with two complete flat bands to a trivial insulator with two dispersive Bloch bands when the hopping strengths $W<J$. As the strength of the disordered hopping further increases, a topological phase transition takes place and the system enters into the topological inverse Anderson insulator phase with two zero-energy edge modes, where the bulk states are extended, even in the presence of strong disorder. Thus, the conventional disordered SSH chain with Bernoulli distribution disordered hopping reveals the topological inverse Anderson insulator phase.

\subsection{Diamond chain}\label{Diamond}
\begin{figure}[tb]
\centering
\includegraphics[width=\columnwidth]{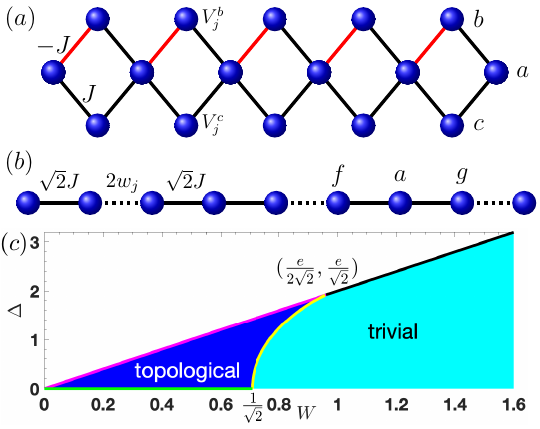}
\caption{$(a)$ Sketch of the disordered $\pi$-flux diamond chain with on-site disorder potentials $V^b_j$ and $V^c_j$. $(b)$ Mapping of the disordered $\pi$-flux diamond chain to a disordered trimerized chain. $(c)$ The phase diagram for the disordered $\pi$-flux diamond chain with the neighboring hopping $J=1$.}
\label{Fig5}
\end{figure}

The next concrete example is the disordered $\pi$-flux diamond chain (see Fig.\ref{Fig5}a)\cite{LonghiS21OL,LiH22PRL, WangHT22PRB, GligoricG20PRA, AhmedA22PRB}, whose Hamiltonian can be written as
\begin{align}
H_0=&J\sum_j^{N-1}\left[a_{j+1}\left(b_{j+1}^{\dagger}+c_{j+1}^{\dagger}-b_j^{\dagger}+c_j^{\dagger}\right) +H.c.\right]\notag \\ 
&+\sum_j^{N}V^b_j b_{j}^{\dagger}b_{j}+V^c_jc_j^{\dagger}c_{j},
\end{align}
where $a_j^{\dagger}$, $b_j^{\dagger}$ and $c_j^{\dagger}$ are fermionic creation operators at the lattice sites $a$, $b$, and $c$ in the $j$-th unit cell. The parameter $J$ denotes the hopping strengths between neighboring sites. $V^b_j$ and $V^c_j$ are the on-site disorder potentials at the lattice sites $b$ and $c$. We also choose the antisymmetric-correlated disorder $V^c_j=-V^b_j$ case.

In the clean limit, there are three complete flat bands at energies $E=0,\pm 2J$.  The compact localized eigenstates of the three flat bands are given by $[b_j^{\dagger}+c_j^{\dagger}\pm 2a_{j+1}^{\dagger}-b_{j+1}^{\dagger}+c_{j+1}^{\dagger}]|0\rangle$ and $[b_j^{\dagger}+c_j^{\dagger}+b_{j+1}^{\dagger}-c_{j+1}^{\dagger}]|0\rangle$. The localization and topological properties of the clean $\pi$-flux diamond chain have been intensely investigated\cite{VidalJ98PRL, KhomerikiR16PRL, KremerM20NTC, PelegriG19PRA, PelegrG19PRA}, known as square-root topological insulators. The three flat bands have the quantized Berry (Zak) phase $\pi$ for the $E=0$ band and the non-quantized Berry phases $\pi/2$ for the two $E=\pm 2J$ bands. The two in-gap topological edge states appear at energies $E=\pm \sqrt2J$ in a finite system under the OBC.

The above analysis for the Creutz ladder can also be used to describe the topological and transport properties of this disordered $\pi$-flux diamond chain. Applying the transformation of the following operators and defining $w_j=(V^c_j-V^b_j)/2$, we can get
\begin{equation}
\left(\begin{array}{c}
f_j^{\dagger} \\
g_j^{\dagger}
\end{array}\right)=\frac{1}{\sqrt{2}}\left(\begin{array}{cc}
-1 & 1 \\
1 & 1
\end{array}\right)\left(\begin{array}{l}
b_j^{\dagger} \\
c_j^{\dagger}
\end{array}\right),
\end{equation}
\begin{equation}
H_{tri}=\sqrt{2}J\sum_j^{N-1}a_{j+1}\left(f_j^{\dagger}+g_{j+1}^{\dagger}\right)+\sum_j^{N}2w_jf_j^{\dagger}g_j+H.c..
\end{equation}

Thus, the disordered $\pi$-flux diamond chain is mapped into a disordered trimerized chain with inter-cell coupling $\sqrt{2}J$ and intra-cell couplings $\sqrt{2}J$ and $2w_j$, as shown in Fig.\ref{Fig5} (a-b).  As argued in the Creutz ladder case, for Bernoulli distribution disorder, we can obtain the energy spectrum equation $E^3-4E(J^2+W^2)+8J^2W^2\cos k=0$, where $k$ is the Bloch wavenumber. The roots of this eigenvalues equation form three dispersive Bloch bands. This disordered system reduces to the disorder-free traditional trimerized chain, whose topological properties have been identified\cite{MartinezAVM19PRA, AnastasiadisA22PRB}. In consequence, for the disordered $\pi$-flux diamond chain,  the system evolves from the square-root topological insulator phase at clean limit into the square-root topological inverse Anderson insulator phase (marked by the green line in Fig.\ref{Fig5}c) in the Bernoulli distribution disorder case. For the uniformly distributed disorder ($\Delta=2W$), the system enters the square-root topological Anderson insulator phase (highlighted by the pink line in Fig.\ref{Fig5}c). For the disordered trimerized chain, the winding number in real space and the localization length of the topological edge states can be used to characterize the topological phases\cite{YaoY21PRA}. Numerical calculations show that the energies of the topological edge states are fixed at $\pm\sqrt{2}J$. The phase diagram for the disordered $\pi$-flux diamond chain with $J=1$ is shown in Fig.\ref{Fig5}c, whose phase structure is similar to that of the disordered $\pi$-flux Creutz ladder. The localization-delocalization (inverse Anderson) transition phenomena of the $\pi$-flux diamond chain with antisymmetric-correlated disorder has been theoretically investigated\cite{LonghiS21OL} and experimentally confirmed\cite{LiH22PRL, WangHT22PRB}.

\section{Conclusions and Discussions}\label{Conclusions}
In short, we uncover a different type of disordered topological insulator phase termed the topological inverse Anderson insulator in several all-band-flat models including the disordered $\pi$-flux Creutz ladder, the fully dimerized SSH chain, and the $\pi$-flux diamond chain. Unlike the topological Anderson insulator, the topological inverse Anderson insulator possesses disorder-induced extended bulk states and topological edge states. By way of the topological invariant, the behaviors of the localization length of the zero-energy modes, and quantum transport, we theoretically demonstrate the existence of this phase. Furthermore, these phenomena can be realized in current experimental techniques such as ultracold atoms\cite{LiH22PRL}, and topoelectrical circuits\cite{WangHT22PRB, ZhangWX23PRL}. This work opens another direction in the search for topological quantum matter, where the disordered systems are topological insulators with extended bulk states and topological edge states.

\section*{Acknowledgements}

This work was supported by the National Natural Science Foundation of China (Grant No. 12074101) and the Natural Science Foundation of Henan (Grant No. 212300410040).

%\bibliographystyle{apsrev4-2}
%\bibliography{TopologicalInsulators}

\end{document}